\documentclass[12 pt]{article}

\usepackage[table]{xcolor}
\usepackage[colorlinks,linkcolor=blue,citecolor=blue,urlcolor=blue]{hyperref}
\usepackage{graphicx} 
\usepackage{amsmath, amsthm, amssymb, lineno}
\usepackage[boxed]{algorithm2e}
\usepackage{cleveref} 
\usepackage{subcaption} 

\usepackage{colortbl}
\usepackage{array}

\usepackage[numbers]{natbib}
\newtheorem{theorem}{Theorem}[section]

\newtheorem{ex}{Example}

\title{Routing functions for parameter space decomposition 
\\ to describe stability landscapes 
of ecological models}

\author{
Joseph Cummings\thanks{Department of Applied and 
Computational Mathematics and Statistics, University of Notre Dame, Notre Dame, IN 46556 (jcummin7@nd.edu)} \and 
Kyle J.-M. Dahlin\thanks{Department of Mathematics and Center for the Mathematics of Biosystems, Virginia Tech, Blacksburg, VA 24061 (kyledahlin@vt.edu)} \and 
Elizabeth Gross \thanks{Department of Mathematics, University of Hawai`i at M\={a}noa,
Honolulu, HI 96822 (egross@hawaii.edu)} \and 
Jonathan D. Hauenstein\thanks{Department of Applied and 
Computational Mathematics and Statistics, University of Notre Dame, Notre Dame, IN 46556 (hauenstein@nd.edu, \url{www.nd.edu/~jhauenst})}}
\date{\today}

\begin{document}

\maketitle

\begin{abstract}
\noindent Changes in environmental or system parameters often drive major biological transitions, including ecosystem collapse, disease outbreaks, and tumor development. Analyzing the stability of steady states in dynamical systems provides critical insight into these transitions. This paper introduces an algebraic framework for analyzing the stability landscapes of ecological models defined by systems of first-order autonomous ordinary differential equations with polynomial or rational rate functions. Using tools from real algebraic geometry, we characterize parameter regions associated with steady-state feasibility and stability via three key boundaries: singular, stability (Routh-Hurwitz), and coordinate boundaries. With these boundaries in mind, we employ routing functions to compute the connected components of parameter space in which the number and type of stable steady states remain constant, revealing the stability landscape of these ecological models. As case studies, we revisit the classical Levins-Culver competition-colonization model and a recent model of coral-bacteria symbioses. In the latter, our method uncovers complex stability regimes, including regions supporting limit cycles, that are inaccessible via traditional techniques. These results demonstrate the potential of our approach to inform ecological theory and intervention strategies in systems with nonlinear interactions and multiple stable states.
\end{abstract}

\section{Introduction}
Biological transitions, while ubiquitous in nature, are often exacerbated or caused by human activities. 
Many of the most consequential events experienced by people or societies are biological transitions -- including tumor development, disease outbreaks, and ecosystem collapse. 
A greater understanding of the causes of these transitions allows us to better detect, prevent, manage, or revert them. 
Table \ref{tab:transitions} provides some examples of biological systems, transitions that they may undergo, and possible causes. 

\begin{table}[]
    \centering
    \begin{tabular}{lllll}
        System     & Initial state   & Altered state & Possible cause  \\\hline
        Cells      & Healthy         & Diseased      & Shift in mutation rates \\
        Cells      & Diseased        & Healthy       & Medication \\
        Drylands   & Semi-arid       & Desert        & Increased grazing pressure \\
        Lake       & Healthy         & Eutrophic     & Increased temperatures \\
        Population & Disease-free    & Epidemic      & Introduction of infected individual \\
        Population & Endemic disease & Disease-free  & Vaccination \\
        Population & Abundant        & Extinct       &  Increased mortality due to predation \\
        Population & Sparse          & Abundant      & Translocation intervention \\\hline
    \end{tabular}
    \caption{Transitions of biological systems and their possible causes }
    \label{tab:transitions}
\end{table}

Transitions occur due to changes to key aspects of these systems (possibly small in magnitude and duration), leading to large-magnitude, long-term changes in the system. 
These transitions may be due to perturbations to system parameters -- for example, eliminating malaria by decreasing the reproduction rate of mosquitoes -- or by short-term changes in the system state caused by external forces -- such as one-time immigration of infected individuals causing a disease outbreak. 
Mathematically, these transitions can be described by assessing the \textit{stability} of the equilibrium states of the system. 
In order to study system transitions, we must first model the system in its current state -- often represented as a steady-state (though no real system is ever truly at equilibrium). 
This is generally done by mechanistically describing the processes that cause changes in the system state as functions of the current system state. 
However, determining the existence and stability of steady states may sometimes be a non-trivial mathematical challenge even when the functions take the form of relatively simple polynomials or rational functions. 
A major goal is to determine how the \textit{stability landscape} shifts as parameter values are perturbed. 
These parameter perturbations can correspond to interventions aimed at curtailing transitions or enabling them in order to achieve a desired state. 

In this paper, we set out to apply a recent algebraic geometry method to determine the stability landscape of any dynamical system described by a system of first-order autonomous ordinary differential equations, $\dot{x} = f\left(x\right)$ where the rate function~$f$ is polynomial or rational in $x$. 
Models of this form are ubiquitous in the mathematical biology literature and often used to describe population dynamics, the spread of disease within a population, or the spread of infection among a population of cells. 
With the use of \textit{routing functions}~\cite{Cummings2024}, we are able to decompose parameter space into (not necessarily disjoint) connected components within which certain equilibria exist and are stable. 
Furthermore, regions of overlap among the connected components indicate where multistability is possible, that is, where perturbations to the system state (not the parameter values), can lead to~transitions. 

As a case study, we consider a model representing the dynamics of a host and two of its symbionts that is described by a system of ODEs with rate functions given by multivariate polynomials of degree two. 
While these functions are simple, the full stability analysis of this model is intractable employing commonly-used techniques. 
We demonstrate using routing functions to analyze slices of the stability landscape
when the coral is intrinsically viable and when the coral is intrinsically non-viable.
As a by-product of this analysis, we also determine the existence
of a region of parameter space wherein the model must exhibit limit cycles 
simply by determining the complement of the identified connected components. 
Our hope is that this new method can be useful for mathematical biologists analyzing similarly formulated systems. 

The rest of the article is organized as follows.
Section~\ref{sec:SteadyState} summarizes steady-state analysis and describes the proposed approach using routing functions. 
This approach is applied to the Levins-Culver model~\cite{Levins1971-lz} in Section~\ref{sec:LevinsCulver} to reproduce known results. 
Section~\ref{sec:Tripartite} applies the approach to a tripartite symbiosis system. 
A short conclusion is provided in Section~\ref{sec:Conclusion}.

\section{Steady-state analysis}\label{sec:SteadyState}

This paper is concerned with understanding the steady states of a dynamical system with polynomial or rational rate function. 
To this end, consider the dynamical system
$$\dot{x} = f(x;a)$$
\noindent where $a = (a_1, \ldots, a_k) \in \mathbb R_{>0}^k$ are parameters, $x=(x_1, \ldots, x_n) \in \mathbb R_{\geq 0}^n$ are variables and $f = (f_1, \ldots, f_n)$ are real-valued polynomial or rational functions. 
For a given $a$, the point~$x$ is a \emph{steady-state} of the system if $f(x;a) = 0$. 
A steady-state $x$ is \emph{locally asymptotically stable}, or simply \emph{stable}, if the eigenvalues of the Jacobian matrix of $f$ with respect to the variables evaluated at $x$, denoted $J_xf(x;a)$, all have negative real parts. 
Otherwise, $x$ is~\emph{unstable}.

Each point $a$ in the parameter space will give rise to a certain number of real positive steady-states and a certain number of stable steady-states. 
In fact, the parameter space~$\mathbb R _{>0}^k$ can be decomposed into regions where these counts are constant. For a polynomial dynamical system as described above, the boundaries of these regions are determined by an algebraic hypersurface (possibly reducible) called the \emph{discriminant locus} (described in Section~\ref{subsec:boundaries}), and the regions themselves are semi-algebraic sets. 
Thus, determining these regions is fundamentally a real algebraic geometry problem. 
See, e.g.,~\cite{RealDiscriminant,GKZ,LRdiscriminant} for more details.

Determining these regions of parameter space are of much interest in mathematical modeling. 
Regions of the parameter space that give rise to multiple real-positive steady-states are often referred to as \emph{regions of multistationarity}, while regions of the parameter space that give rise to multiple real-positive stable steady-states are often referred to as \emph{regions of multistability}.
A concrete understanding of these regions can aid in model selection and discrimination \cite{feliu2013simplifying}.  
In some cases, just determining whether regions of multistationarity or regions of multistability 
exist for a dynamical system can be a challenging question~\cite{conradi2017identifying,joshi2015survey-cp}. 

As an illustration of the challenge to determine these regions of parameter space,
consider the following model that represents a predator population $y$ with a foraging rate that follows a Holling Type II functional response and a prey population $x$ with a strong Allee effect:
\begin{equation}\label{eq:PP_Allee}
\begin{aligned}
\frac{dx}{dt} &= x\left(1 - \frac{x}{K}\right)\left(\frac{x}{A} - 1\right) - \frac{\alpha xy}{1 + \beta x}, \\
\frac{dy}{dt} &= y\left(-m + \frac{\gamma x}{1 + \beta x}\right).
\end{aligned}
\end{equation}
The parameters of this system are $\alpha$, $\beta$, $\gamma > 0$. 
The right-hand side is rational where the denominators do not vanish on the sets of interest. 
In fact, there are four steady-states:  
total extinction $(0,0)$, 
two prey-only steady-states, $\left(A,0\right)$ and $\left(K,0\right)$, and a coexistence steady-state,~$(x^{*}, y^{*})$, provided that 
$x^{*} = \frac{m}{\gamma - m\beta}>0$
and 
\begin{equation} \label{eq:two-dim-coexistence}
y^* = \frac{(1+\beta x^*)(K-x^*)(x^*-A)}{\alpha A K} > 0 ~~\Longrightarrow~~
(K-x^*)(x^*-A)>0.
\end{equation}
Note that total extinction is stable with eigenvalues $-1$ and $-m$. 
Assume that 
$$0<K<x^*=\frac{m}{\ensuremath{\ensuremath{\gamma}}-m\beta}<A.$$
Then, $\left(A,0\right)$ is unstable and $\left(K,0\right)$ is stable. 
Since the system is only two-dimensional, the stability of the coexistence steady-state $(x^{*}, y^{*})$ can be determined simply by computing the signs of the trace and determinant of the Jacobian matrix evaluated at $(x^{*}, y^{*})$, say $J^*$:
\begin{equation*}
    \begin{aligned}
        \text{tr}(J^*)&=\left(1-\frac{x^{*}}{K}\right)\left(\frac{x^{*}}{A}-1\right)\left(\frac{1+2\beta x^{*}}{1+\beta x^{*}}\right)+\left(1-\frac{x^{*}}{K}\frac{x^{*}}{A}\right),\\
        \det(J^*)&=\left(\frac{\gamma}{\left(1+\beta x^{*}\right)^{2}}\right)x^{*}\left(1-\frac{x^{*}}{K}\right)\left(\frac{x^{*}}{A}-1\right).
    \end{aligned}
\end{equation*}
The positivity of $\det(J^*)$ is straightforward
via $x^*>0$ and \eqref{eq:two-dim-coexistence}.  
Thus, the coexistence steady-state is stable
if $\text{tr}(J^*)<0$
and the corresponding stability boundary 
arises~from~\mbox{$\text{tr}(J^*)=0$}.

If we fix all parameters except one, we can use $\text{tr}(J^*)$ to straightforwardly determine the stability of the coexistence steady-state. Suppose that \mbox{$(\alpha,\beta,\gamma,m,K) = (2,3/2,1,1/2,1)$}. 
Now, for example,~if  $A= 9/4$, then $(x^*,y^*) = (2,2/9)$ is stable with
$\text{tr}(J^*)=-7/12<0$ and $\det(J^*)=1/72>0$. 
When $A = 30/11$, then $(x^*,y^*)=(2,8/15)$ 
has $\text{tr}(J^*)=0$ and thus lies
on the boundary of stability.  
Furthermore, as one increases $A$ to pass through
 the boundary, for example at $A = 3$,
then $(x^*,y^*)=(2,2/3)$ and 
$\text{tr}(J^*)=1/4>0$
so that~$(x^*,y^*)$ is unstable. 
But extending this analysis to higher dimensions, say by allowing $K$ to also be a free parameter, leads to a much more difficult problem.

In this work, which is demonstrated using two models from ecology, we use techniques based on routing functions (described in Section~\ref{subsec:routing-functions}) to compute the decomposition of the parameter space. 
A full description of the decomposition of the parameter space is called a \emph{landscape}, or \emph{paramter landscape}.  
We will focus primarily on the \emph{stability landscape} for our systems of interest.
In the stability landscapes described below, the number and type (e.g., coexistence) of the stable steady-states are constant in each region.

 
 
 








\subsection{Boundaries}
\label{subsec:boundaries}

As mentioned above, for a polynomial or rational dynamical system, the region boundaries of a parameter landscape are contained in an algebraic hypersurface. 
The following describes how to obtain this hypersurface in general.  
To aid the description, we use language from computational algebraic geometry, e.g., see \cite{CLO} for a general background. 
For simplicity, we make some assumptions about the system $f$ that hold throughout. 
Given a generic $a^*$, we assume that the system $f(x;a^*) = 0$ has a finite number of isolated solutions, say $d$, over the complex numbers, all of which are nonsingular, i.e., $f(x;a^*)=0$ implies \mbox{$\det(J_xf(x;a^*))\neq 0$}.
Moreover, we assume that, for all $a$, the system $f(x;a)=0$ has exactly $d$ solutions~counting~multiplicity.  

In considering how to compute the region boundaries, we need to first consider the different ways that the number of real positive stable steady-states can change as we vary the parameters $a$. 
For our application, this will result in three types of algebraic boundaries: the \emph{singular boundary}, the \emph{stability (Routh-Hurwitz) boundary}, and \emph{coordinate boundaries}. 
The singular boundary, which is the classical discriminant locus, arises from parameters $a$ where there exists $x$ with $f(x;a) = 0$ and $\det(J_xf(x;a))=0$.
The stability (Routh-Hurwitz) boundary arises when the greatest real part among all eigenvalues (the spectral abscissa of $J_xf(x;a))$, becomes $0$.
Finally, a coordinate boundary arises when a coordinate of a steady-state becomes $0$.

\paragraph{Singular boundary} 
The first way that the number of real positive stable steady-states can change is by the \emph{total} number of real solutions changing. 
This can only occur when $f(x;a)=0$ has a singular solution, i.e., $\det(J_xf(x;a))=0$.
The (Euclidean) closure of the set of such parameters~$a$, called the \emph{singular boundary}, forms an algebraic subset of the parameter space. 
For an illustrative example, consider the quadratic equation $$f(x;a)=a_1 x^2 + a_2 x + a_3 = 0.$$
Thus, the singular boundary arises when $$f(x;a)=a_1x^2 + a_2x + a_3 = 0 ~~\hbox{and}~~ f'(x;a) = 2a_1 x + a_2 = 0$$ which, upon eliminating $x$, yields the classical discriminant locus defined by
$$D(a) = a_2^2 - 4a_1a_3 =0.$$
In particular, $f(x;a)=0$ has 
two real solutions when $D(a)>0$,
no real solutions when $D(a)<0$, 
and one real solution (with multiplicity two) when $D(a)=0$.  

As illustrated in the quadratic above, 
the singular boundary arises as the zero set of
an \emph{elimination ideal}, which can be computed algorithmically using Gr\"{o}bner bases. 
Since elimination ideals are key to this perspective computationally, we will present the boundaries, including the singular boundary, 
through an ideal-theoretic perspective. 
The first piece is the \emph{equilibrium ideal}
$\mathcal{I}_f = \langle f \rangle \subset \mathbb{C}[x,a]$.  The corresponding algebraic set
is the \emph{equilibrium~variety} 
$$\mathcal V_f = \mathcal V(\mathcal I_f) = \{ (x, a) \in \mathbb C^{n+k}~|~f(x;a) = 0\}.$$  
Notice that both the parameters $a$ and the variables $x$ are being treated as indeterminates in this setup.
The projection map $\pi_a: \mathbb C^{n+k} \to \mathbb C^{k}$ defined by $\pi_a(x,a)=a$ 
is the canonical projection onto the parameter space.
The geometric analog to 
the algebraic operation of elimination is projection.

With the assumptions above, 
the \emph{ideal of the singular boundary} 
is defined by
$$\left(\mathcal I_f + \langle \det(J_xf)\rangle\right)\cap\mathbb{C}[a]$$
and the \emph{singular boundary} is the corresponding
algebraic set. 
For example, via Gr\"obner bases, 
there is an algorithmic approach to compute
the ideal of the singular boundary.

Especially for biological and ecological systems,
the equilibrium ideal $\mathcal I_f$ need not be
prime, i.e., the equilibrium variety $\mathcal V_f$
is reducible.  
For example, the system in \eqref{eq:PP_Allee} 
has four prime components.  
When $\mathcal I_f$ is not prime,
one can construct the ideal of the singular
boundary for each prime component, 
which is typically an easier computation.

\paragraph{Stability (Routh-Hurwitz) boundary} 
The second way that the number of real positive stable steady-states can change is by a change in the \emph{stability} of a steady-state,
i.e., when the real part of an eigenvalue
becomes~$0$.   
Since we will use the Routh-Hurwitz criterion \cite{Hurwitz}
to determine stability, we will refer to this boundary as the \emph{Routh-Hurwitz boundary}.
Since the singular boundary arises
from an eigenvalue becoming $0$,
the \emph{Routh-Hurwitz boundary}
contains the singular boundary.

Given parameters $a$, let $x$
be a corresponding steady-state.
A necessary Routh-Hur\-witz 
condition for~$x$ to be stable
is that the coefficients of the characteristic polynomial of~$J_xf(x;a)$,
namely, $\det(\lambda I - J_xf(x;a))$,
are all positive.  
In particular, when $f$ is polynomial or rational,
there are polynomials or rational functions $c_n(x;a),c_{n-1}(x;a),\dots,c_0(x;a)$
with
$$\det(\lambda I - J_xf(x;a))
= c_n(x;a)\lambda^n + c_{n-1}(x;a) \lambda^{n-1}
+ \cdots + c_0(x;a).$$
Note that $c_n(x;a) = 1$,
$c_{n-1}(x;a) = -\text{tr}(J_xf(x;a))$,
and $c_0(x;a) = (-1)^n \det(J_xf(x;z))$.
Hence, the necessary condition is that 
$c_i(x;a) > 0$ for $i=0,\dots,n$.

With $c_n(x;a)=1$, the 
sufficient Routh-Hurwitz condition 
can be stated as all entries of the first column
of the so-called Routh array are all positive.  
Each entry of the Routh array is a rational function
in the coefficients $c_i(x;a)$ 
of the characteristic polynomial,
with the denominator arising from another entry
so only the numerators need to be considered.
We call each numerator a \emph{Routh polynomial}.
A full description of the Routh array can be found in, for example, \cite{dorf2011modern}.  

Since the systems in this paper either 
have $n=2$ or $n=3$, we explicitly provide the corresponding conditions where $c_n(x;a)=1$. 
For $n=2$, the Routh-Hurwitz criterion 
for stability is simply that the two Routh
polynomials, namely $\beta_y(x;a)$ and $c_0(x;a)$,
are positive.  This corresponds with
$\text{tr}(J_xf(x;a)) < 0$
and $\det(J_xf(x;a)) > 0$ as utilized in the preliminary example at the start of the section.

For $n=3$, the Routh-Hurwitz criterion 
for stability is that the following three Routh polynomials are~positive:
$$\beta_z(x;a),~~
\beta_z(x;a)\beta_y(x;a) - c_3(x;a)c_0(x;a),
~~c_0(x;a).$$
With $c_3(x;a)=1$, positivity of 
these three Routh polynomials imply $\beta_y(x;a)>0$.

The \emph{Routh-Hurwitz boundary} is
the (Euclidean) closure of the set of parameters
$a$ such that $f(x;a)=0$ and some Routh polynomial
vanishes.  With $c_n(x;a)=1$, there are 
$n$ Routh polynomials, say $r_1(x;a), \dots, r_n(x;a)$.  Thus, for $R(x;a) = \prod_{i=1}^n r_i(x;a)$, the \emph{ideal of the Routh-Hurwitz boundary} is defined by 
$$\left(\mathcal I_f + \langle R\rangle\right)\cap\mathbb{C}[a].$$
As with the singular boundary, one can
construct the ideal of the Routh-Hurwitz boundary
for each prime component of $\mathcal I_f$
and each Routh polynomial.

\paragraph{Coordinate boundary}
Finally, the third way that the number of real positive stable steady-states can change is if a steady-state is no longer positive.  
Given a variable, say~$x_j$, 
the \emph{$x_j$-boundary} is the Zariski closure of the set of all parameter values where the system $f(x;a)=0$ has a solution with $x_j=0$.  Similar, to the singular boundary and the Routh-Hurwitz boundary, the \emph{ideal of the $x_j$-boundary} is
$$\left(\mathcal I_f + \langle x_j\rangle\right)\cap\mathbb{C}[a].$$

As in the illustrative example~\eqref{eq:PP_Allee} 
and as will
be seen in Sections \ref{sec:LevinsCulver}
and~\ref{sec:Tripartite}, it could be the case that 
certain types of steady-states have some coordinates identically
equal to zero, e.g., extinction 
and prey-only in the illustrative example above.  
In such cases, the corresponding ideal for the 
coordinate boundaries would be the 
zero ideal, $\langle 0\rangle$, since every
parameter value~$a$ has a solution with that coordinate equal to zero.
Thus, to obtain relevant information associated with 
coordinate boundaries, one needs to consider each prime
component of $\mathcal I_f$ separately and 
assess which coordinate boundaries are relevant
for each prime component.  

\bigskip

Putting everything together and removing redundancies,
one obtains the \emph{total boundary}.
The framework described above is flexible in that
one can consider the total boundary for the entire system
or only for certain types of steady-states.
Nonetheless, with the assumptions above, the total boundary 
is either empty or is a union of hypersurfaces, called
a \emph{hypersurface arrangement}, in $\mathbb{C}^k$.
Since an empty total boundary is not interesting,
we assume that the total boundary, $\mathcal{B}\subset\mathbb{C}^k$, 
is a hypersurface arrangement of the form
\begin{equation}\label{eq:Boundary}
\mathcal{B} = \bigcup_{i=1}^m \{a~|~g_i(a)=0\}
\end{equation}
where $g_1,\dots,g_m$ are nonconstant polynomials in $a$.  
In particular, the type and number of 
stable steady-states remain constant 
on each connected component of $\mathbb{R}_{>0}^k\setminus\mathcal{B}$.
The last step is to compute such connected components,
which is considered next.



\subsection{Routing function and connected components}
\label{subsec:routing-functions}

The following describes using \emph{routing functions},
e.g., see~\cite{Cummings2024,Hong2010,HRSS2020},
to compute the connected components 
of $\mathbb{R}_{>0}^k\setminus\mathcal{B}$,
where $\mathcal{B}$ is as in \eqref{eq:Boundary}.
For a full accounting of routing functions and 
justification for the algorithm proposed in this section, 
we refer the reader to \cite{Cummings2024}.
From \eqref{eq:Boundary}, one obtains
a rational function 
\begin{equation}\label{eq:RoutingFunctionForm}
    r_{c}(a) = \frac{g_1(a)\cdots g_m(a)}{(1 + 
    (a_1-\beta_y)^2+\cdots+(a_k-c_k)^2)^D}
\end{equation}
where $c\in\mathbb{R}_{>0}^k$ and $2D > \sum_{i=1}^m \deg(g_i)$. 
For a generic choice of $c \in \mathbb{R}_{>0}^k$, 
the function~$r_{c}$ satisfies the following conditions \cite[Thm.~3.4]{Cummings2024}:
\begin{enumerate}
    \item For all $\epsilon > 0$, there exists $\delta>0$ so that if $\|x\|>\delta$, then $|r_{c}(x)| < \epsilon$.
    \item There are finitely many critical points of $r_{c}$,
    i.e., solutions of $\nabla r_{c}(a) = 0$,
    on $\mathbb{R}_{>0}^k \setminus\mathcal{B}$.
    \item For each critical point $a \in \mathbb{R}_{>0}^k \setminus \mathcal{B}$ of $r_{c}$, the Hessian of $r_{c}$ evaluated at $a$, namely $Hr_{c}(a)$, 
    is invertible, i.e., each critical point is non-degenerate.
    \item For each $\alpha \in \mathbb{R}\setminus\{0\}$, there is at most one critical point $a$ of $r_{c}$ satisfying $r_{c}(a) = \alpha$.
\end{enumerate}
Such a function $r_{c}$ is called a \emph{routing function} for~$\mathbb{R}_{>0}^k \setminus \mathcal{B}$ and the critical points of $r_{c}$ in~$\mathbb{R}_{>0}^k \setminus \mathcal{B}$ are called \emph{routing points}. 
The \emph{index} of a routing point $a$ is the number of eigenvalues of $Hr_{c}(a)$ which have the same sign as~$r_{c}(a)$. 
Hence, the index 0 routing points are precisely the local maxima when~$r_{c}>0$ and local minima when~$r_{c}<0$.
An eigenvector $v$ of $Hr_{c}(a)$ of unit length
with a corresponding eigenvalue $\lambda$ that has the same sign as~$r_{c}(a)$ is 
said to be an \emph{unstable eigenvector}
with $v$ and $-v$ being the corresponding \emph{unstable
eigenvector directions}.

One outcome of Theorem~4.4 of \cite{Cummings2024} 
is that connected components of 
$\mathbb{R}^k_{>0}\setminus\mathcal{B}$ 
are in one-to-one 
correspondence with the connected sets of routing points 
where the connections arise from gradient ascent/descent
along unstable eigenvector directions as follows.
If $a$ is a routing 
point with index $>$ 0 and $v$ is an unstable eigenvector 
direction, consider the following initial~value~problem:
\begin{equation}
\label{Eqn:DiffEq}
    \begin{cases}
    \dot{y}(t) = \mathrm{sign}(r_{c}(a)) \cdot \nabla r_{c}(y), \\
    \lim_{t\to 0^+}\frac{\dot{y}(t)}{\|\dot{y}(t)\|} = v, \\
    \lim_{t \to 0^+}y(t) = a.
    \end{cases}
\end{equation}
By \cite[Prop.~4.2]{Cummings2024}, the trajectory $y(t)$
is well-defined and $\lim_{t\rightarrow\infty} y(t)$ must
also be a routing point.  
Looping over all routing points and all 
unstable eigenvector directions
yields the relevant connections between the routing points.

It is natural to construct a graph
whose vertices are the routing points of $r_{c}$
in $\mathbb{R}_{>0}^k\setminus\mathcal B$
and whose edges arise by the connections 
between the routing points via \ref{Eqn:DiffEq}. 
In fact, Algorithm~\ref{alg:Connectivity}
computes such a graph with the following justified by \cite[Thm.~4.4]{Cummings2024}.

\begin{theorem}\label{thm:Graph}
The connected components of the graph $G$ 
computed by Algorithm~\ref{alg:Connectivity}
are in one-to-one correspondence with the
connected components of $\mathbb{R}^k_{>0}\setminus\mathcal{B}$.
\end{theorem}

Since $\mathcal{B}$ is a hypersurface arrangement, 
there is a recently developed
Julia package called \texttt{HypersurfaceRegions.jl} \cite{breiding2024computingarrangementshypersurfaces}
that can be used to perform the computations
in Algorithm~\ref{alg:Connectivity}.

\begin{algorithm}[!t]
\caption{Connected Components}  
\label{alg:Connectivity}
\SetAlgoLined
\KwIn{A routing function $r$ satisfying conditions 1-4 above
associated with $\mathbb{R}_{>0}^k\setminus\mathcal B$.}
\KwOut{A graph $G$ 
whose vertices are the routing points of $r$ and whose
connected components are in one-to-one correspondence with
the connected components of $\mathbb{R}_{>0}^k \setminus \mathcal{B}$.}

Compute the routing points of $r$ on $\mathbb{R}_{>0}^k \setminus \mathcal{B}$, say $z_1,\dots,z_\ell$.

Initialize the graph $G$ whose vertices are $z_1,\dotsc,z_\ell$ 
with no edges.

\For{$j=1,\dots,\ell$}{
  \ForEach{unstable eigenvector $v$ of $H r(z_j)$}{ 
    Compute limit routing point $z_{+}$ starting from $z_j$
    in the direction $v$ with respect to $r$
    using~\eqref{Eqn:DiffEq}.

    Add the edge $(z_j, z_{+})$ to $G$.

    Compute limit routing point $z_{-}$ starting from $z_j$
    in the direction $-v$ with respect to $r$
    using~\eqref{Eqn:DiffEq}.

    Add the edge $(z_j, z_{-})$ to $G$.
  }
}

\Return{G}

\end{algorithm}

\begin{figure}[!t]
    \centering
    \includegraphics[width=0.5\linewidth]{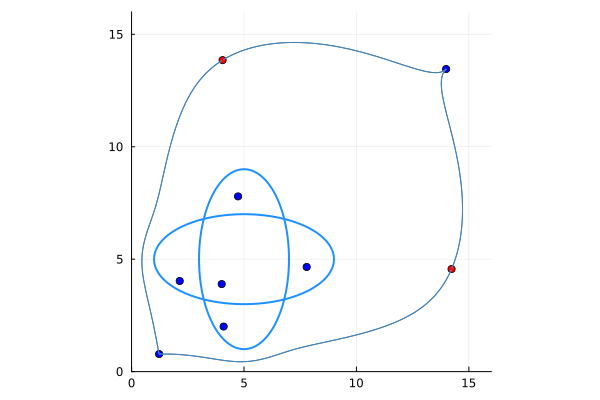}
    \caption{Computing the 6 regions in $\mathbb{R}^2_{>0}\setminus\mathcal{B}$
    where the boundaries are the two ellipses (thick blue) and coordinate axes.
    Routing points (index 0 in blue, index 1 in red) and gradient paths (thin blue)
    are also plotted demonstrating $6$ connected components.}
    \label{fig:planeExample}
\end{figure}

\begin{ex}
Consider computing the connected components of $\mathbb{R}^2_{>0} \setminus \mathcal{B}$
where the hypersurface arrangement $\mathcal{B}$ arises from the two coordinate boundaries
and two ellipses, i.e.,
$$g_1(a) = a_1,~~g_2(a) = a_2,~~g_3(a) = \frac{\left(a_1-5\right)^{2}}{16}+\frac{\left(a_2-5\right)^{2}}{4}-1,~~g_4(a)= \frac{\left(a_1-5\right)^{2}}{4}+\frac{\left(a_2-5\right)^{2}}{16}-1.
$$
For a generic choice of $c\in\mathbb{R}^2_{>0}$,
$r_{c}(a)$ as in \eqref{eq:RoutingFunctionForm} yields a routing function. 
For our experiment, 
we used $c = (1.2, 0.7)$ yielding
9 routing points of $r_{c}$
on $\mathbb{R}^2_{>0} \setminus \mathcal{B}$: $7$ have index 0 and
$2$ have index~1.  Using gradient paths along unstable eigenvector directions
as in Algorithm \ref{alg:Connectivity}, we find that there are 6 connected components of 
$\mathbb{R}^2_{>0} \setminus \mathcal{B}$
as observed in~Figure~\ref{fig:planeExample}.
\end{ex}

Given a point $a\in\mathbb{R}^k_{>0}\setminus \mathcal{B}$
which is not a routing point, 
the output of Algorithm~\ref{alg:Connectivity} can be used to determine
which connected component of $\mathbb{R}^k_{>0}\setminus \mathcal{B}$
contains $a$.  For this, one simply solves the following initial value problem:
\begin{equation}
\label{Eqn:DiffEq2}
    \begin{cases}
    \dot{y}(t) = \mathrm{sign}(r_{c}(a)) \cdot \nabla r_{c}(y), \\
    y(0) = a.
    \end{cases}
\end{equation}
By \cite[Prop.~4.1]{Cummings2024}, the trajectory $y(t)$
is well-defined and $\lim_{t\rightarrow\infty} y(t)$ must
be a routing point in the same connected component as $a$.  

The computationally-intensive parts of this approach 
are computing the total boundary, 
computing the routing points, and tracking gradient ascent/descent paths.
As the size of the system increases, these parts become challenging 
on both symbolic and numerical computations.  Reducing the number
of parameters by fixing some to constant values is a common approach
to reduce the computational burden to obtain some insights.

\section{Levins and Culver model of competition-colonization trade-off}\label{sec:LevinsCulver}

Scientists have long sought to explain how coexistence can arise through spatial competition among species. While the full answer certainly involves several axes of variation \citep{Amarasekare2003-ht}, many theorists have focused on individual mechanisms to improve our understanding of coexistence. 
One simple theory suggests that two species can coexist through a competition-colonization trade-off wherein one species is superior at colonizing unoccupied space while the other species is capable of displacing the former from occupied patches \citep{Levins1971-lz}.

The Levins-Culver model exhibits such a competition-colonization trade-off for two populations occupying the same niche and can be summarized by the system of equations:
\begin{equation}
    \begin{aligned}\label{eq:Levins-Culver}
            \frac{dy}{dt} &= \beta_y y \left(1-y\right) - \gamma_y y, \\
            \frac{dz}{dt} &= \beta_z z\left(1 - y - z\right) - \beta_y y z - \gamma_z z ,
    \end{aligned}
\end{equation}
where $y$ is the fraction of the niche occupied by species $1$ and $z$ is the fraction of the niche occupied by species $2$ \citep{Yu2001-fo}. The colonization rates of species $1$ and $2$ are limited by the availability of resources, represented by the terms $1-y$ and $1-y-z$, respectively. However, species $1$ is able to colonize any area currently occupied by species $2$. 

At first glance, it may seem that species $2$ will always be excluded from the niche through replacement by species $1$. Through a stability analysis, it can be shown that if the intrinsic colonization rate of species $2$, $\beta_z$, is large enough, then the two species can coexist in the long term. 
More precisely, if $\beta_z > \beta_y \left(\beta_y + \gamma_z - \gamma_y\right)/\gamma_y$ and $\beta_y > \gamma_y$, then the coexistence equilibrium,
which satisfies
\begin{equation}
    \begin{aligned}\label{eq:LC-equilibria}
        y &= 1-\frac{\gamma_{y}}{\beta_{y}}, \\
        z &= \frac{\gamma_{y}}{\beta_{y}}-\frac{1}{\beta_{z}}\left(\gamma_{z}+\beta_{y}-\gamma_{y}\right),
    \end{aligned}
\end{equation}
is locally asymptotically stable.



To finish this section, we give a proof that if $\beta_z > \beta_y \left(\beta_y + \gamma_z - \gamma_y\right)/\gamma_y$ and $\beta_y > \gamma_y$, then there is a locally asymptotically stable coexistence equilibrium using routing functions.
The idea is to first construct the region boundaries in parameter space as described in Section~\ref{subsec:boundaries}, and then it is enough to check there is a locally asymptotically stable coexistence equilibrium at a single routing point in each region.

First, we construct the boundaries. The ideal $\mathcal{I}_f$ in this case is determined from the right-hand sides of equation \eqref{eq:Levins-Culver}
\[
    \langle \beta_y y \left(1-y\right) - \gamma_y y, \beta_z z \left(1 - y - z\right) - \beta_y y z - \gamma_z z  \rangle. 
\]
The ideal is not prime; hence, we consider a primary decomposition which takes the form $\mathcal{I}_f = \bigcap_{i=1}^4 \mathcal{J}_f^{(i)}$ where 
\begin{align*}
    \mathcal{J}_f^{(1)} &= \langle y, z\rangle \\
    \mathcal{J}_f^{(2)} &= \left\langle \beta_y\left(1-y\right)-\gamma_y,z\right\rangle \\
    \mathcal{J}_f^{(3)} &= \left\langle y,\beta_z\left(1-z\right)-\gamma_z\right\rangle  \\
    \mathcal{J}_f^{(4)} &= \left\langle \beta_y\left(1-y\right)-\gamma_y,\beta_z\left(1-y-z\right)-\beta_y+\gamma_z\right\rangle 
\end{align*}
The characteristic polynomial of the system is of the form $\lambda^2 + a_1\lambda + a_0$ where $a_0$ and $a_1$ are below.
\begin{align*}
    a_0 &= 2y^{2}\beta_y^{2}+2y^{2}\beta_y\beta_z+4yz\beta_y\beta_z - y\beta_y^{2}-3y\beta_y\beta_z \\
    &\;\;\;\;\,-2z\beta_y\beta_z+y\beta_y\gamma_y+y\beta_z\gamma_y+2z\beta_z\gamma_y \\
    &\;\;\;\;\,+2y\beta_y\gamma_z+\beta_y\beta_z-\beta_z\gamma_y-\beta_y\gamma_z+\gamma_y\gamma_z\\
    a_1 &= 3y\beta_y+y\beta_z+2z\beta_z-\beta_y-\beta_z+\gamma_y+\gamma_z
\end{align*}
In order to find equations for the region boundaries, we need to compute the following elimination ideals and intersect them:
\begin{flalign*}
    (\mathcal{J}_f^{(1)} \cap \langle a_0 a_1\rangle) \cap \mathbb{C}[\beta_y,\beta_z,\gamma_y,\gamma_z], \\
    (\mathcal{J}_f^{(2)} \cap \langle y a_0 a_1\rangle) \cap \mathbb{C}[\beta_y,\beta_z,\gamma_y,\gamma_z], \\
    (\mathcal{J}_f^{(3)} \cap \langle z a_0 a_1\rangle) \cap \mathbb{C}[\beta_y,\beta_z,\gamma_y,\gamma_z], \\
    (\mathcal{J}_f^{(4)} \cap \langle y z a_0 a_1\rangle) \cap \mathbb{C}[\beta_y,\beta_z,\gamma_y,\gamma_z].
\end{flalign*}
The resulting ideal is principal with the total boundary 
defined by the vanishing of the following polynomials:
\[\begin{array}{l}
g_1 = \beta_y-\gamma_y, ~~~~ g_2 = \beta_z-\gamma_z, ~~~~ g_3=\beta_y-\beta_z-\gamma_y+\gamma_z, ~~~~ g_4 = \beta_y+\beta_z-\gamma_y-\gamma_z, \\
g_5 = \beta_z\gamma_y-\beta_y\gamma_z, ~~~ g_6 = \beta_y^{2}-\beta_y\gamma_y-\beta_z\gamma_y+\beta_y\gamma_z, ~~~
g_7 = 2\,\beta_y^{2}-2\,\beta_y\gamma_y-\beta_z\gamma_y+\beta_y\gamma_z.
\end{array}\]
Note that $g_1$ is the boundary of $\beta_y>\gamma_y$ and $g_6$ 
is the boundary of $\beta_z > \beta_y(\beta_y+\gamma_z-\gamma_y)/\gamma_y$.  
Since we are only interested in parameter values that are positive, we include $\beta_y,\beta_z,\gamma_y,$ and $\gamma_z$ as boundaries along with $g_1,\dots,g_7$ above, so the resulting arrangement $\mathcal{A}$ consists of 11 hypersurfaces. 
Using the Julia package \texttt{HypersurfaceRegions.jl}, we find that there are 168 regions in total; however, there are only 16 regions that divide up the positive orthant~$\mathbb{R}_{>0}^4$. These are the 16 regions we need to test.

Of these 16 regions, there are 2 that make up the region where  $\beta_z > \beta_y \left(\beta_y + \gamma_z - \gamma_y\right)/\gamma_y$ and $\beta_y > \gamma_y$. We can pick a representative routing point from each of these regions and check to see if these parameter values give rise to a locally asymptotically stable coexistence equilibrium. The parameter values are 
\begin{align*}
    (\beta_y,\beta_z,\gamma_y,\gamma_z) &= (1.03941, 1.76600, 0.93685, 0.22883),\\
    (\beta_y,\beta_z,\gamma_y,\gamma_z) &= (1.21871, 1.48986, 0.75558, 0.23395).
\end{align*}
It is straightforward to check that each gives rise to a locally asymptotically stable coexistence equilibrium; therefore, every point in these regions gives rise to a stable equilibrium as this property can only change by crossing a boundary. On the other hand, if we check any of the other 14 regions, we find that these parameter values never give rise to a locally asymptotically stable coexistence equilibrium confirming the result stated earlier.

\section{Tripartite symbiosis}\label{sec:Tripartite}
In a recent article, Gibbs et al., formulated a model of coral-bacteria symbioses by extending the Levins-Culver model \citep{Gibbs2024-gx}. 
In their model, the patch landscape, in this case, the coral reef available for bacteria to colonize, is no longer fixed but dynamically changing over time. 
In addition, one type of bacteria is assumed to increase the reproduction rate of the coral it colonizes (\textit{mutualism}), while the other type of bacteria increases coral mortality (\textit{parasitism}). 

The coral populations are assumed to inhabit a reef with a total carrying capacity scaled to one. 
In the absence of bacteria, the unoccupied host coral population, $x$, grows over bare reef at the intrinsic per-capita growth rate $b$ and dies off at the mortality rate $d$. 
The growth of the coral is inhibited by density dependence so that its size cannot exceed the carrying capacity of the bare reef. 

The state variables $y$ and $z$ correspond to the amount of coral occupied by parasitic and mutualistic bacteria, respectively. These parasitic and mutualistic bacteria populations colonize coral through mass-action interaction with contact rates $\beta_y$ and $\beta_z$, respectively. 
The two types of bacteria cannot occupy the same patch of coral, hence $x+y+z\le 1$. 
Both bacterial types are obligate symbionts of the coral host, so their carrying capacities are limited by the available coral area. 
Additionally, if the coral they occupy dies off (which occurs at rate $d$), the bacteria die off as well. 
Due to virulence, the coral occupied by parasitic bacteria endures additional mortality at the rate $\tilde{d}$, which we refer to as the ``death detriment'' for brevity. 
Bacteria also die off on their own, leaving the coral it previously inhabited unoccupied, at rates $\gamma_y$ or $\gamma_z$. 
The total mortality rate for parasitic bacteria is, therefore, $d + \tilde{d} +\gamma_y$. 
Parasitic bacteria also colonize coral that is already occupied by mutualistic bacteria at the per-capita rate $\beta_y y$, thereby limiting the carrying capacity of mutualistic bacteria to $1-x-y$. 
Finally, colonization by the mutualistic bacteria induces an increased growth rate in the coral, $\tilde{b}$, which we refer to as ``birth benefit'' for brevity. 

\begin{figure}[t!]
    \centering
    \includegraphics[width=.8 \linewidth, trim = {0cm 7.5cm 5cm 0cm},clip]{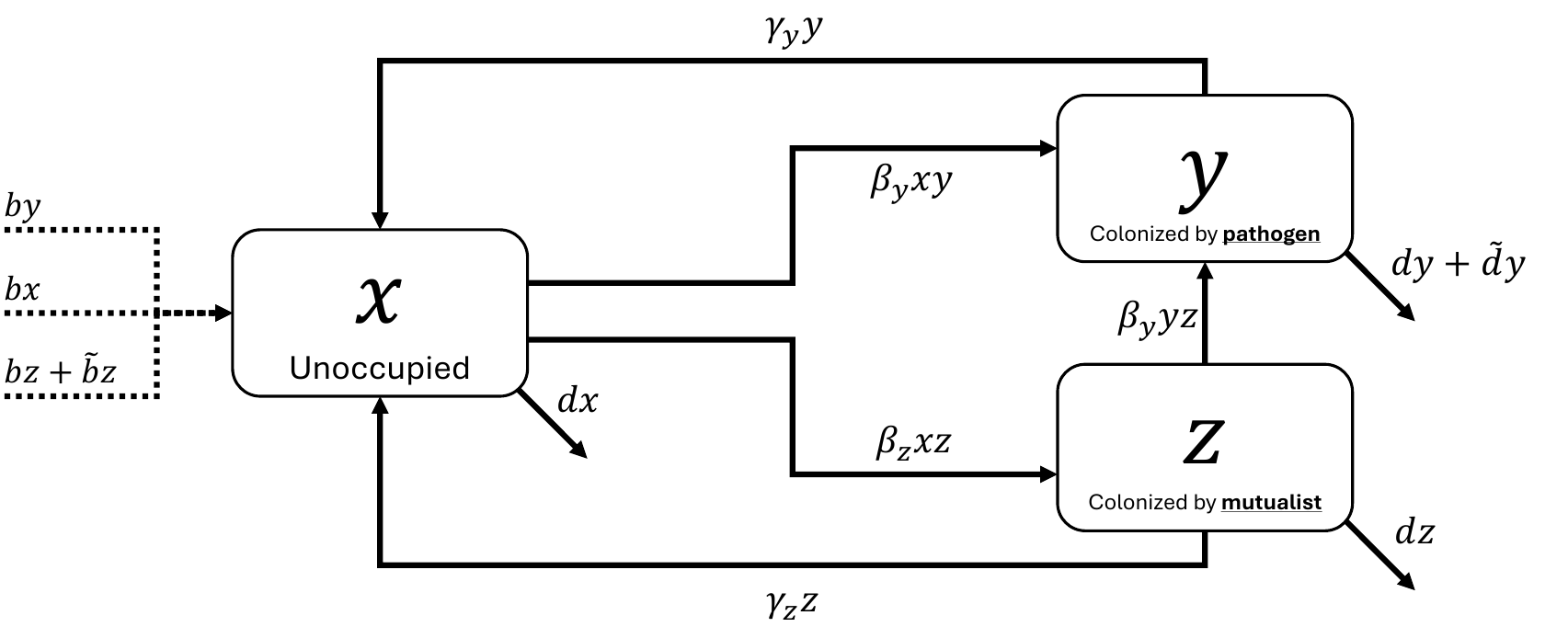}
    \caption{The coral-bacteria symbioses compartmental model represented by system \eqref{eq:xyz-ODEs}. }
    \label{fig:diagram}
\end{figure}

The full system of ordinary differential equations describing this model is given by equations~\eqref{eq:xyz-ODEs} and illustrated in Figure \ref{fig:diagram}:
\begin{equation}
    \begin{aligned}\label{eq:xyz-ODEs}
    \frac{dx}{dt}&=[b(x+y)+(b+\tilde{b})z]\left[1-\left(x+y+z\right)\right]-\beta_{y}xy-\beta_{z}xz+\gamma_{y}y+\gamma_{z}z-dx,\\
    \frac{dy}{dt}&=\beta_{y}xy+\beta_{y}yz-(d+\tilde{d}+\gamma_{y})y,\\
    \frac{dz}{dt}&=\beta_{z}xz-\beta_{y}yz-\left(d+\gamma_{z}\right)z.
    \end{aligned}
\end{equation}

Of interest are the criteria (i.e., sets of parameters and initial conditions) for which i) coexistence among all three populations is stable; ii) mutualistic or parasitic bacteria are excluded from the system in the long term; iii) mutualistic or parasitic bacteria are able to invade a system where it is initially absent. 
Studying the criteria for coexistence or exclusion provides insights into the key drivers for the assembly of stable tripartite symbioses with competing symbionts. 
Such knowledge could inform the development of conservation strategies for coral reefs affected by acidifying and warming oceans by providing the criteria in which beneficial bacteria stably coexist with coral, with or without the presence of~harmful~bacteria.

We characterize the eight possible equilibria of system \eqref{eq:xyz-ODEs} into five types: 
\begin{itemize}
    \item total extinction equilibrium $E_0 = \left(0,0,0\right)$; 
    \item bacteria-free equilibria $E_x = \left(+,0,0\right)$;
    \item mutualistic bacteria exclusion equilibria~$E_{xy} = \left(+,+,0\right)$; 
    \item parasitic bacteria exclusion equilibria  $E_{xz} = \left(+,0,+\right)$; and
    \item full coexistence equilibria $E_{\textrm{co}} = \left(+,+,+\right)$. 
\end{itemize}
The results of \cite{Gibbs2024-gx} determined criteria for the existence and stability of these equilibria under the assumption that the mutualistic bacteria had no effect on the growth of the coral ($\tilde{b}=0$) as well as when both bacteria had no effect whatsoever on coral life history ($\tilde{b}=0$ and $\tilde{d}=0$). 
The other cases were considered numerically, where it was discovered that stable limit cycles around the coexistence equilibrium point could occur for certain parameter combinations. 
However, a full description of the stability landscape when $\tilde{b}$ and $\tilde{d}$ were both positive was~not~provided. 

We originally sought to fully describe the equilibrium dynamics of \eqref{eq:xyz-ODEs} by utilizing the techniques outlined in Section~\ref{subsec:routing-functions} just as 
was accomplished for for the Levins-Culver model in Section~\ref{sec:LevinsCulver}. 
However, due to the large the number of parameters, 
the Gr{\"o}bner basis computations for computing the boundaries 
did not finish for the entire model.  
Thus, to reduce to number of parameters, 
we set all the death rates equal to 1 and the birth rate of the parasitic bacteria is set to 5, i.e. $d = \gamma_y = \gamma_z = 1$ and $\beta_y = 5$. 
The resulting model has four free parameters: $b, \beta_z, \tilde{b},$~and~$\tilde{d}$.

Proceeding as in the end of Section~\ref{sec:LevinsCulver}, we compute the equations defining the total boundary. It turns out that the ideal $\mathcal{I}_f \subset \mathbb{C}[x,y,z,b,\beta_z,\tilde{b},\tilde{d}]$ in this case has 5 primary components, one corresponding to each of the five equilibria state types: $E_0, E_x, E_{xy}, E_{xz}, E_\textrm{co}$. 
After adding in the Routh-Hurwitz conditions and eliminating $x,y,z$, we are left with a principal ideal $\langle g \rangle \subset \mathbb{C}[b,\beta_z,\tilde{b},\tilde{d}]$. The polynomial $g$ has 31 distinct irreducible factors, $g_1,\dotsc,g_{31}$, whose degrees range from 1 to 20.  In fact, there are 6 linear factors,
7 quadratic factors,
 2 cubic factors,
4 quartic factors,
4 quintic factors,
3 degree 9 factors,
and the remaining 5 factors have degrees 6, 7, 10, 13, and 20.
We end up with the total boundary $\mathcal{B} = \bigcup_{i=1}^{31} \{a \in \mathbb{R}^4 ~|~ g_i(a) = 0\}$.
These computations were done in {\tt Macaulay2} \cite{M2} and can be found in a \href{https://github.com/jcu237/Routing-functions-for-parameter-space-decomposition-to-describe-stability-landscapes}{GitHub repository}.

The next step is to compute the full set of routing points where each connected component of the complement of the boundary in $\mathbb{R}^4_{>0}$ contains a single routing point. 
Since the degree of the numerator of the routing function, namely $g_1\cdots g_{31}$, has degree $145$, computing the routing points in full generality was numerically unstable.  
Thus, we consider two major cases: when $b = 2$ and when $b = 0.5$. 
These values were chosen because they correspond to cases where the coral can persist on its own ($b=2>d=1$) and where it cannot ($b=0.5<d=1$). 
Correspondingly, $b - 1$ is one of the six linear factors in $g$.
For each of these values of $b$, 
we consider the stability landscape for increasing values of the colonization rate of the mutualistic bacteria: \mbox{$\beta_z = 2.5, 3, 3.5, 3.9, 4.1, 5.9, 6.1$} and $10$. The values $3.9,4.1,5.9$ and $6.1$ were chosen since when $b = 2$, the linear polynomials $\beta_z - 4$ and $\beta_z - 6$ are factors of $g\vert_{b=2}$. 

For each of these 16 slices, we construct a routing function and compute the connected components of $\mathbb{R}^2_{>0} \setminus\mathcal{B}$. A routing point in each connected component was chosen and the possible steady states for these parameter values were computed. Finally, we colored each region according to which steady states were possible at the corresponding routing point. These computations were done using the Julia package \texttt{HomotopyContinuation.jl} \cite{HomotopyContinuation}. The results of these computations can be found in the next section, and all computations are freely available in a \href{https://github.com/jcu237/Routing-functions-for-parameter-space-decomposition-to-describe-stability-landscapes}{GitHub repository}.

\subsection{Simulations}

The following analyzes the three free parameters:
the birth benefit of mutualist bacteria to the coral host~($\tilde{b}$), the death detriment imposed by parasitic bacteria to the coral host~($\tilde{d}$), and the colonization rate of mutualistic bacteria~($\beta_z$). 
To focus on the effects of the symbionts on their host (via $\tilde{b}$ and $\tilde{d}$), our results are presented as ``slices'' along the $\beta_z$ axis. 
The names and assigned colors for each stable steady state set are presented in Table \ref{tab:eq_colors}. 

\begin{table}[b!]
\centering
\caption{Color key for stability landscape labels}
\renewcommand{\arraystretch}{1.5}
\begin{tabular}{|>{\columncolor{white}}c|>{\columncolor{white}}l|>{\columncolor{white}}c|}
\hline
\textbf{Label} & \textbf{Name} & \textbf{Color} \\
\hline
{$E_0 = (0,0,0)$}    & Total extinction & \cellcolor[HTML]{E73F74} \hspace{2cm} \\
{$E_x = (+,0,0)$}    & Coral only & \cellcolor[HTML]{F1CE63} \hspace{2cm} \\
{$E_{xy} = (+,+,0)$}    & Mutualist exclusion & \cellcolor[HTML]{77AADD} \hspace{2cm} \\
{$E_{xz} = (+,0,+)$}    & Pathogen exclusion & \cellcolor[HTML]{9467BD} \hspace{2cm} \\
{$E_{co} = (+,+,+)$}    & Coexistence & \cellcolor[HTML]{AAAA00} \hspace{2cm} \\
{$S_{0,xy} = \left\{E_0, E_{xy}\right\}$}  & Extinction or pathogen exclusion & \cellcolor[HTML]{FF9D9A} \hspace{2cm} \\
{$S_{0,co}=\left\{E_0, E_{co}\right\}$}  & Extinction or coexistence & \cellcolor[HTML]{99DDFF} \hspace{2cm} \\
{$S_{x,xz} = \left\{ E_x, E_{xz}\right\}$}  & Coral only or pathogen exclusion & \cellcolor[HTML]{B07AA1} \hspace{2cm} \\
{$S_{x,co} = \left\{E_x, E_{co}\right\}$}  & Coral only or coexistence & \cellcolor[HTML]{225522} \hspace{2cm} \\
\hline
\end{tabular}
\label{tab:eq_colors}
\end{table}

\subsubsection{\texorpdfstring{Intrinsically viable coral population ($b=2$)}{Intrinsically viable coral population (b=2)}}
\begin{figure}[p]
    \centering
    \scalebox{1.0}{ 
    \begin{minipage}[t]{0.25\textwidth}
        \centering
        \includegraphics[width=\linewidth]{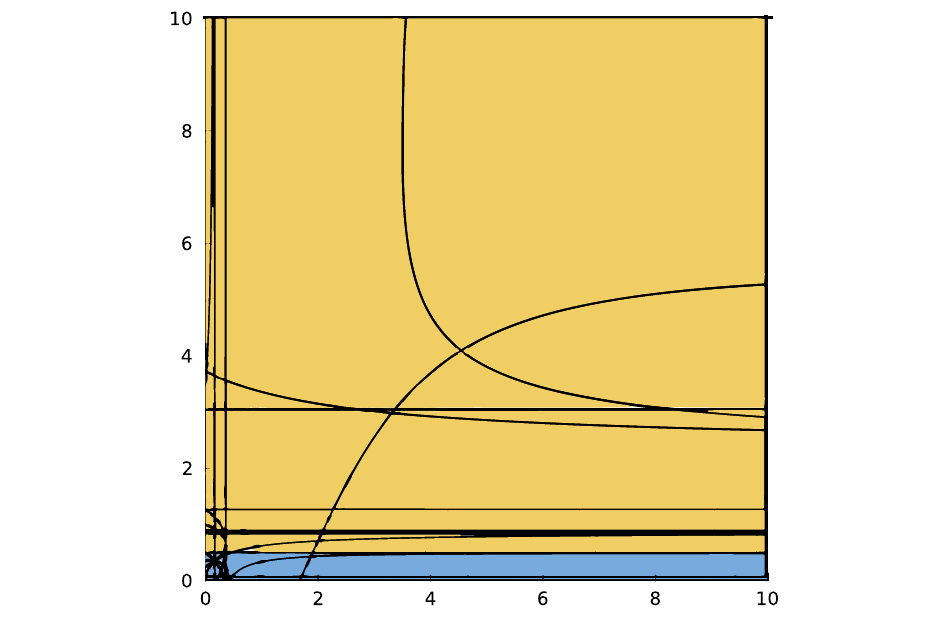}
        \subcaption*{\small (a) $\beta_z = 2.5$}

        \centering
        \includegraphics[width=\linewidth]{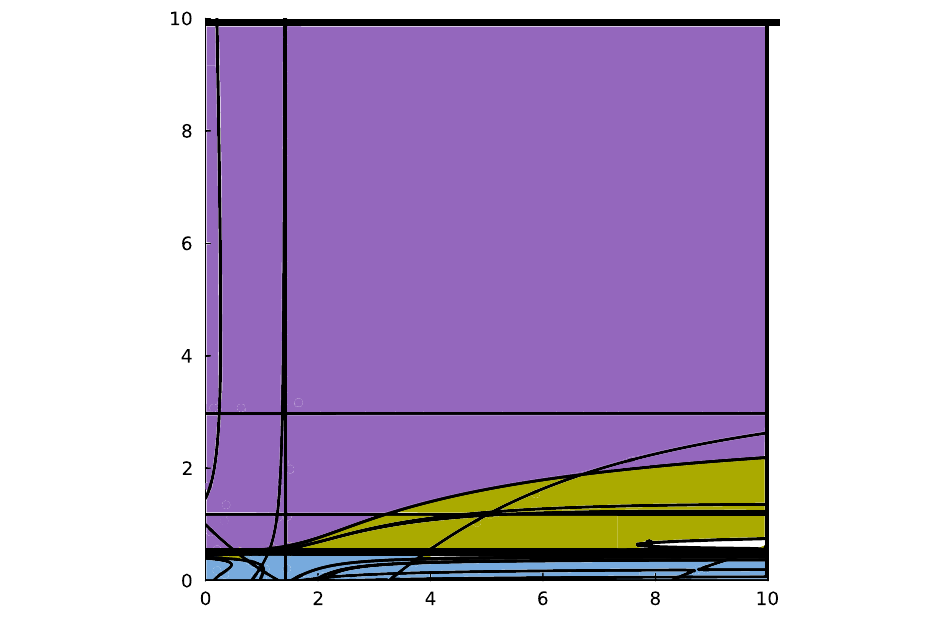}
        \subcaption*{\small (e) $\beta_z = 4.1$}

    \end{minipage}
    
    \begin{minipage}[t]{0.25\textwidth}
        \centering
        \includegraphics[width=\linewidth]{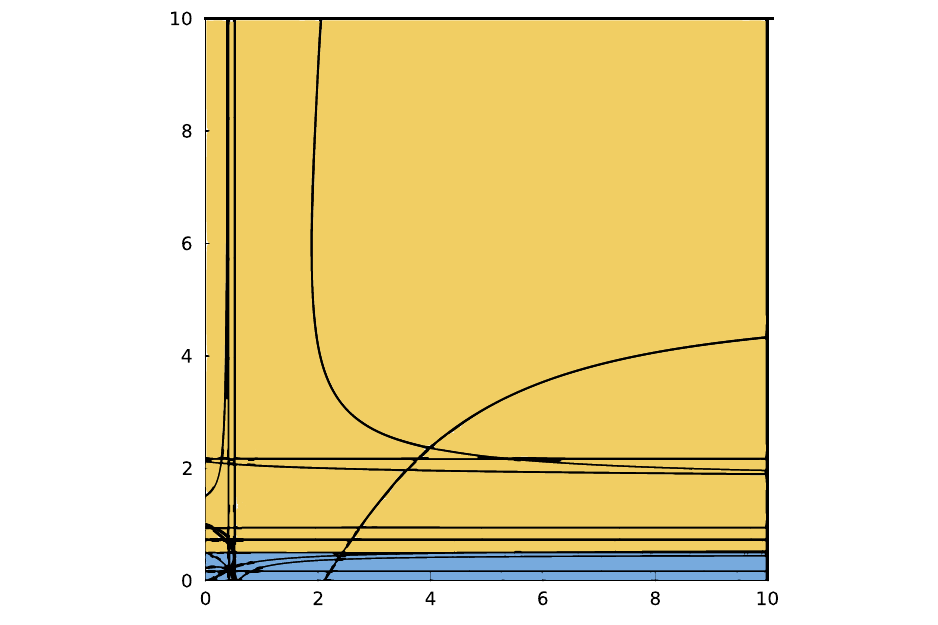}
        \subcaption*{\small (b) $\beta_z = 3$}

        \centering
        \includegraphics[width=\linewidth]{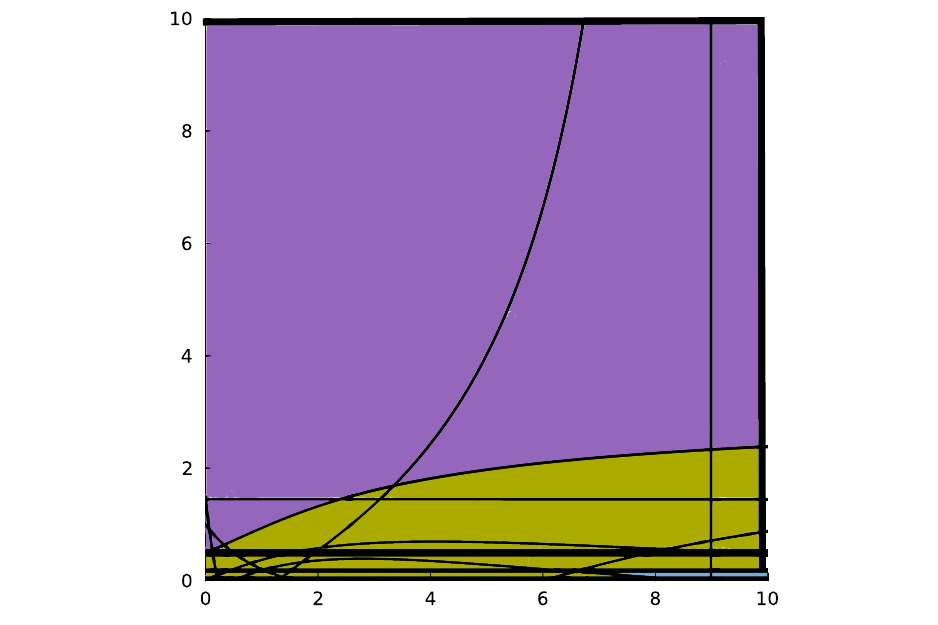}
        \subcaption*{\small (f) $\beta_z = 5.9$}

    \end{minipage}
    
    \begin{minipage}[t]{0.25\textwidth}
        \centering
        \includegraphics[width=\linewidth]{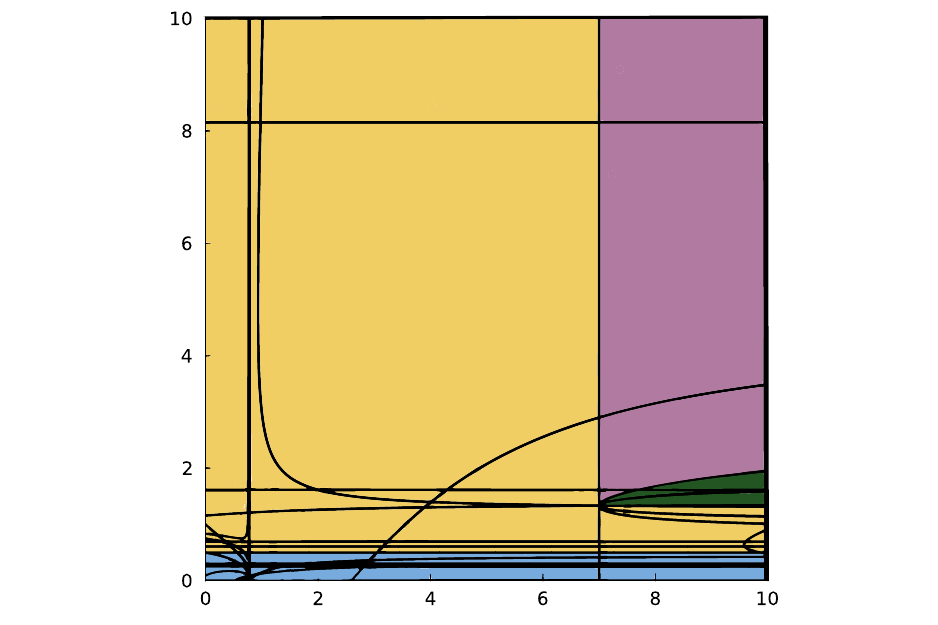}
        \subcaption*{\small (c) $\beta_z = 3.5$}

        \centering
        \includegraphics[width=\linewidth]{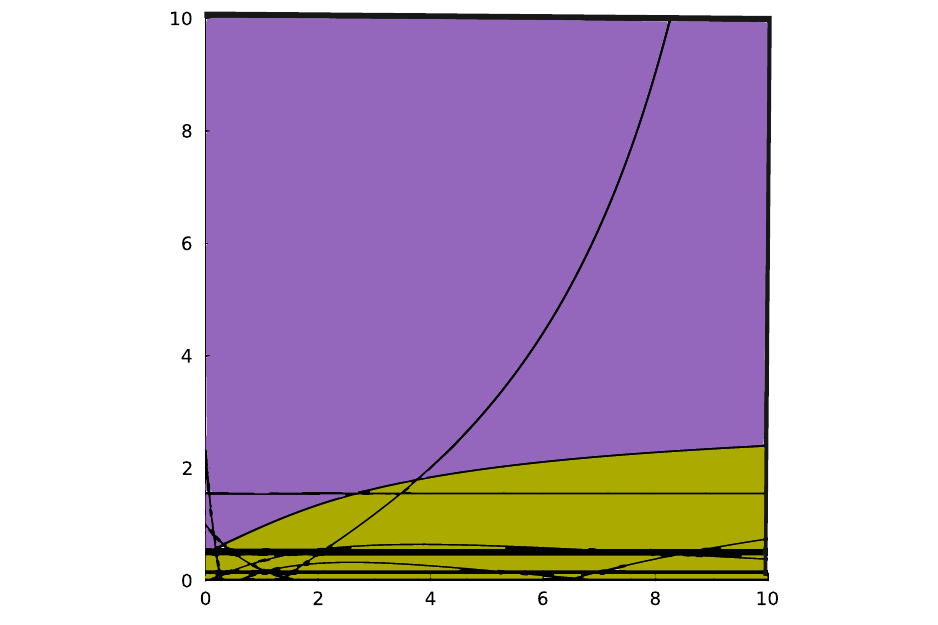}
        \subcaption*{\small (g) $\beta_z = 6.1$}

    \end{minipage}
    
    \begin{minipage}[t]{0.25\textwidth}
        \centering
        \includegraphics[width=\linewidth]{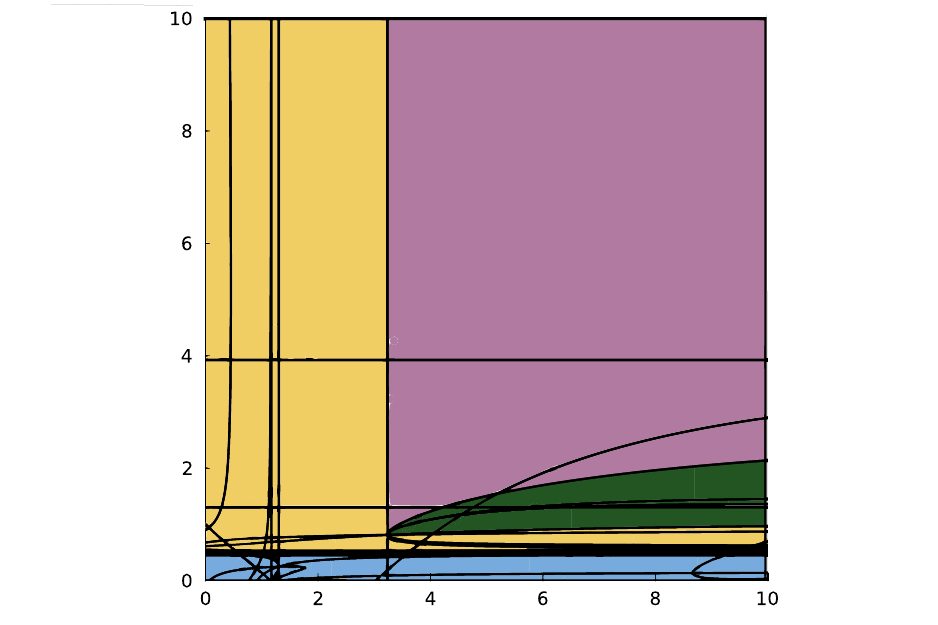}
        \subcaption*{\small (d) $\beta_z = 3.9$}

        \centering
        \includegraphics[width=\linewidth]{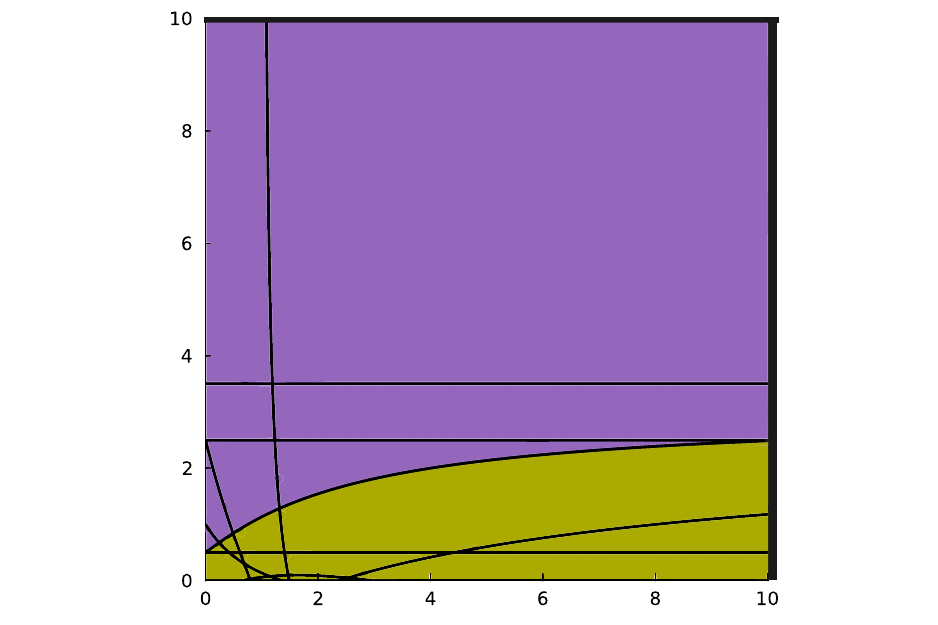}
        \subcaption*{\small (h) $\beta_z = 10$}

    \end{minipage}
    }
    \caption{Stability landscapes for increasing values of $\beta_z$ with $b = 2$. In this case, the intrinsic coral birth rate $b$ exceeds the intrinsic coral mortality rate $d=1$. $\tilde{b}$ values are varied along the $x$-axis and $\tilde{d}$ values along the $y$-axis from zero to ten.}
    \label{fig:bx2}
\end{figure}

Figure \ref{fig:bx2} shows the changes in the $b=2$ stability landscape as $\beta_z$ is increased from 2.5 to~10. 
At low values of $\beta_z$, the coral only steady state $E_x$ is stable, with a very narrow stability region for mutualist exclusion $E_{xy}$ at low levels of death detriment $\tilde{d}$. 
As $\beta_z$ is increased to~3.5, there is bistability for sufficiently large values of reproductive benefit, $\tilde{b}$, and mortality detriment, $\tilde{d}$. 
The mutualist population is positive only within these regions of bistability, 
$S_{x,xz}=\{E_x,E_{xz}\}$ and $S_{x,co}=\{E_x,E_{co}\}$. 
As $\beta_z$ crosses four, the stable sets become $E_{xy}$, $E_{xz}$, and $E_{co}$ and bistability is no longer possible. 
Across most of the landscape, the pathogen is now excluded, with a very narrow region of mutualist exclusion for low $\tilde{d}$, a small region where coexistence is stable, and a very small region where no steady states are stable (shown in white). 
As $\beta_z$ is further increased, the region of mutualistic extinction vanishes, and the coexistence region grows slightly. 

In this case, the extinction state $E_0$ is not stable for any combination of $\beta_z$, $\tilde{b}$, and $\tilde{d}$ since the coral can persist without the assistance of the mutualistic bacteria. 
For our chosen~$\beta_z$ values, bistability only occurred for values of $\beta_z=3.5$ and $3.9$.  When $\beta_z=4.1$, there is a small white region in (e) for which
none of the steady states are stable and the model must have a limit cycle.
For any value of $\beta_z$, a stable steady state can only contain parasitic bacteria if $\tilde{d}$ is sufficiently low, with an upper limit of approximately $\tilde{d} = 2.5$ across all values~of~$\beta_z$. 

\begin{figure}[p]
    \centering
    \scalebox{1.0}{ 
    \begin{minipage}[t]{0.25\textwidth}
        \centering
        \includegraphics[width=\linewidth]{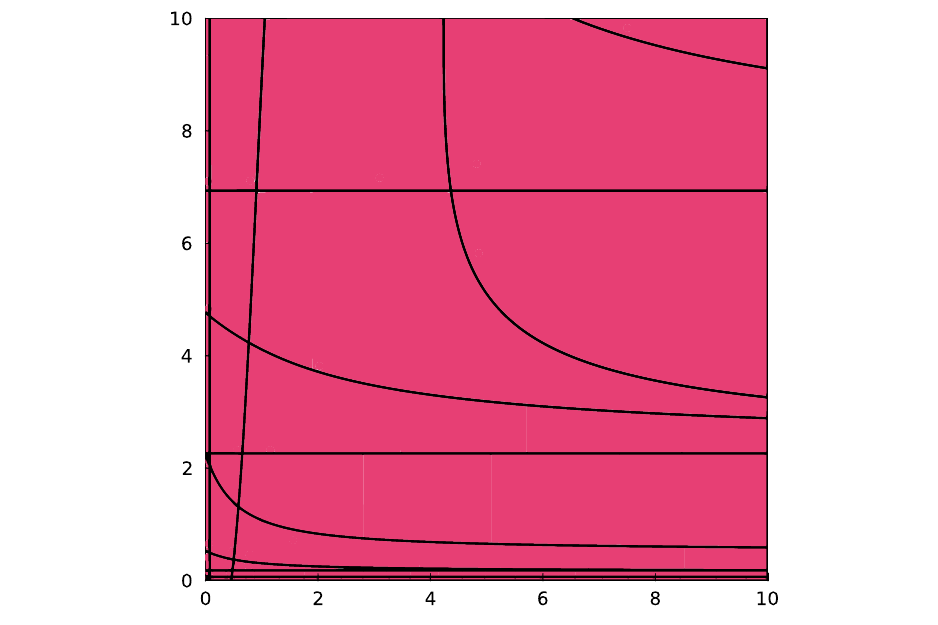}
        \subcaption*{\small (a) $\beta_z = 2.5$}

        \centering
        \includegraphics[width=\linewidth]{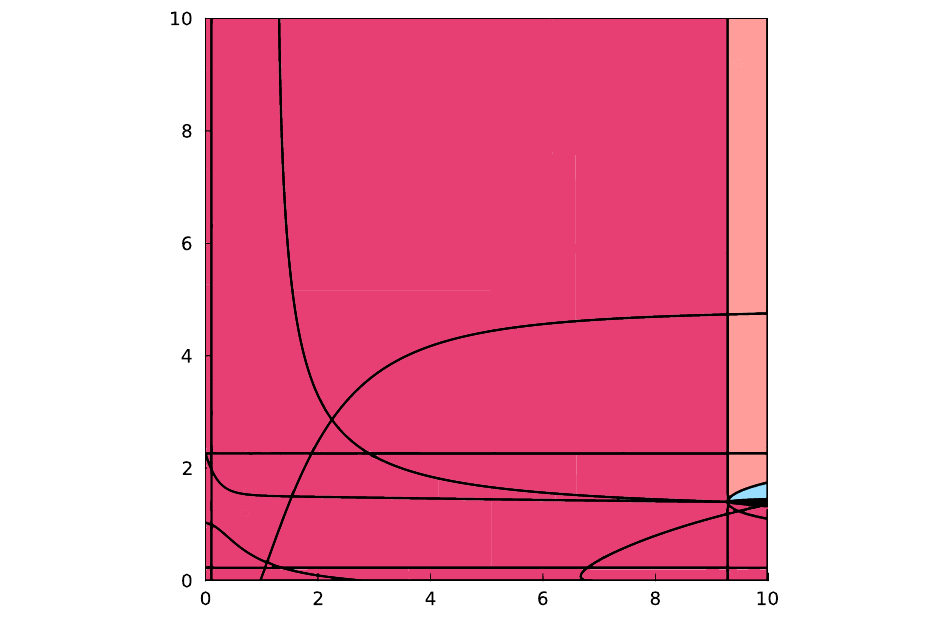}
        \subcaption*{\small (e) $\beta_z = 4.1$}

    \end{minipage}
    
    \begin{minipage}[t]{0.25\textwidth}
        \centering
        \includegraphics[width=\linewidth]{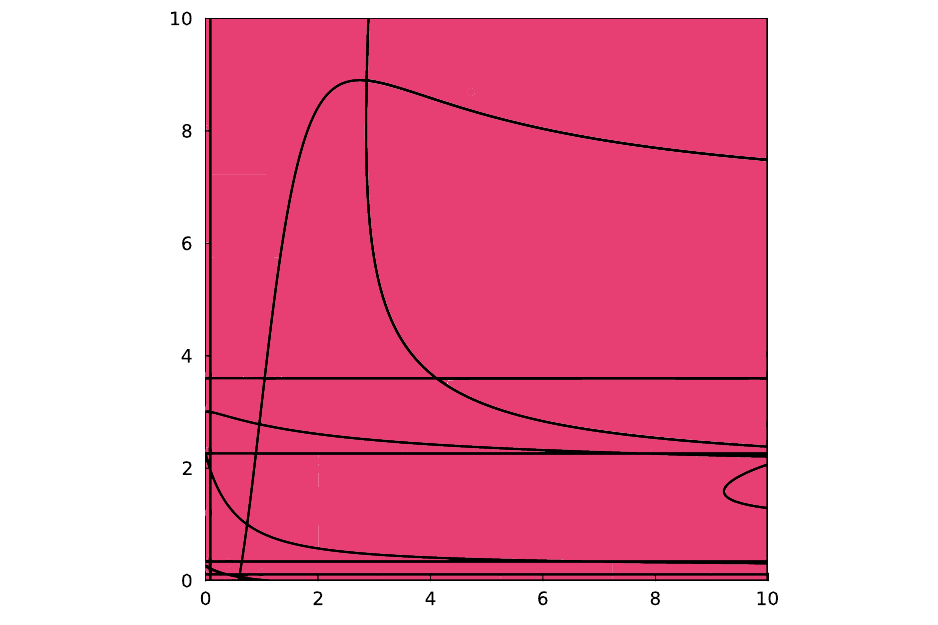}
        \subcaption*{\small (b) $\beta_z = 3$}

        \centering
        \includegraphics[width=\linewidth]{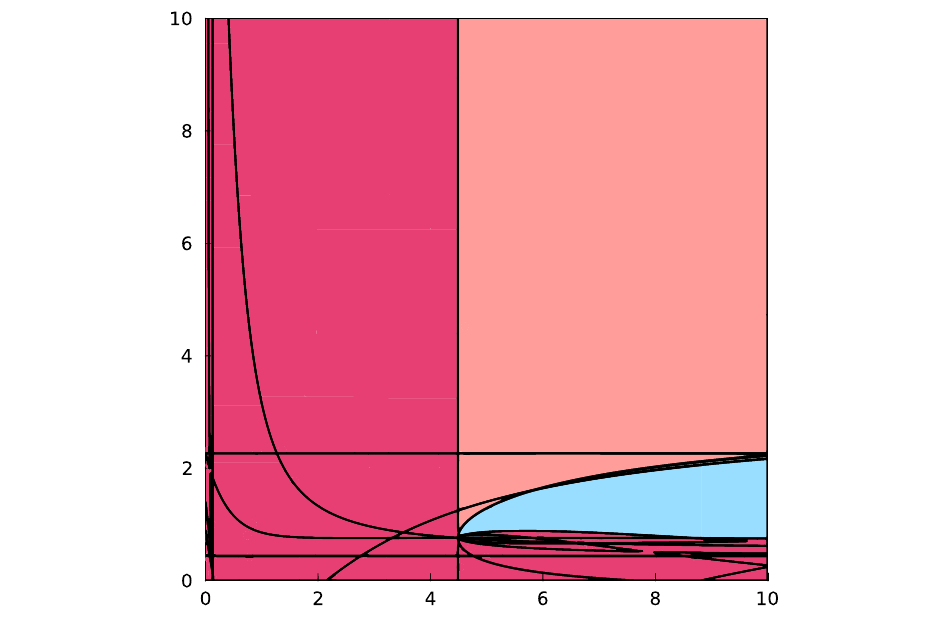}
        \subcaption*{\small (f) $\beta_z = 5.9$}

    \end{minipage}
    
    \begin{minipage}[t]{0.25\textwidth}
        \centering
        \includegraphics[width=\linewidth]{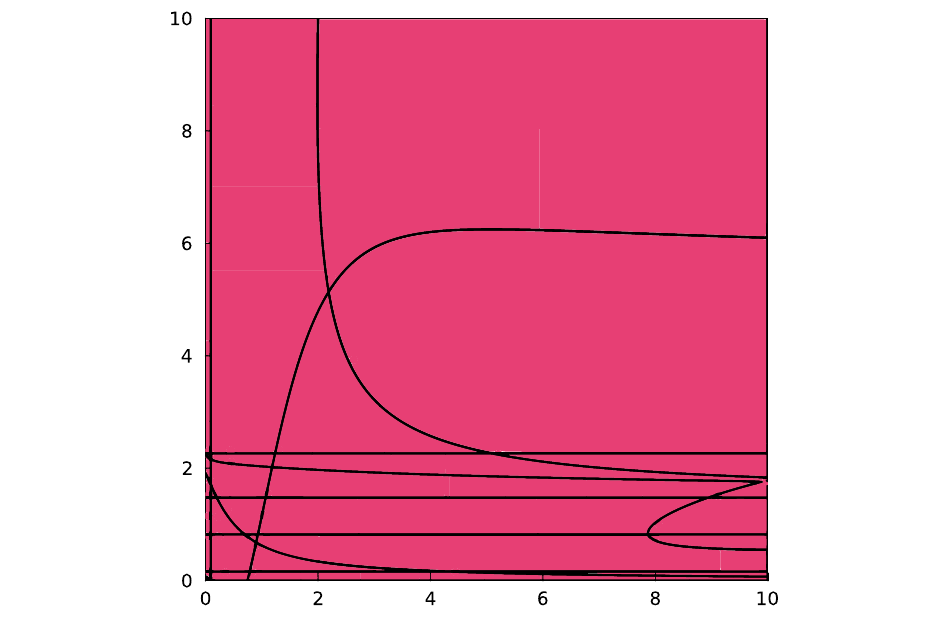}
        \subcaption*{\small (c) $\beta_z = 3.5$}

        \centering
        \includegraphics[width=\linewidth]{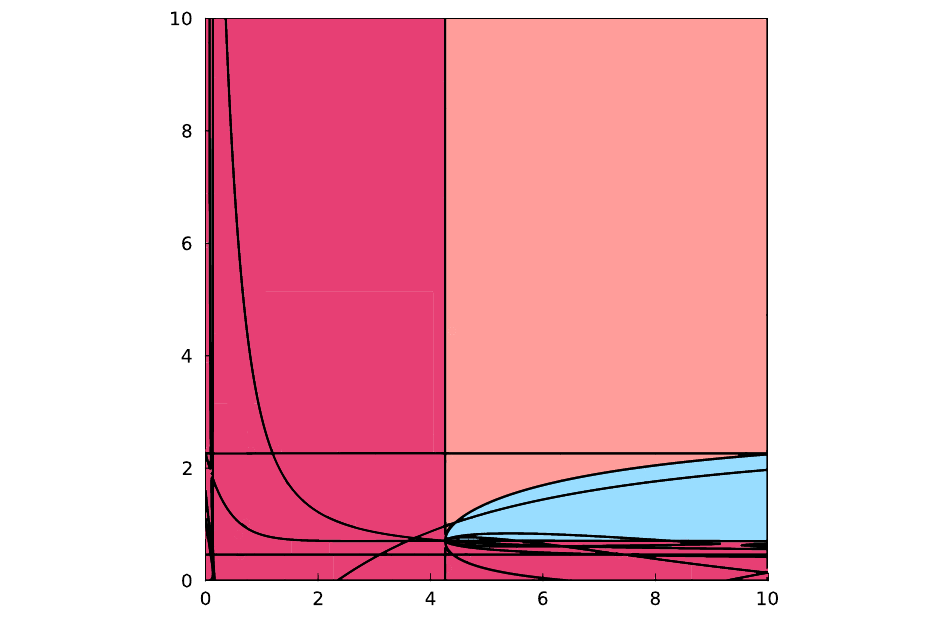}
        \subcaption*{\small (g) $\beta_z = 6.1$}

    \end{minipage}
    
    \begin{minipage}[t]{0.25\textwidth}
        \centering
        \includegraphics[width=\linewidth]{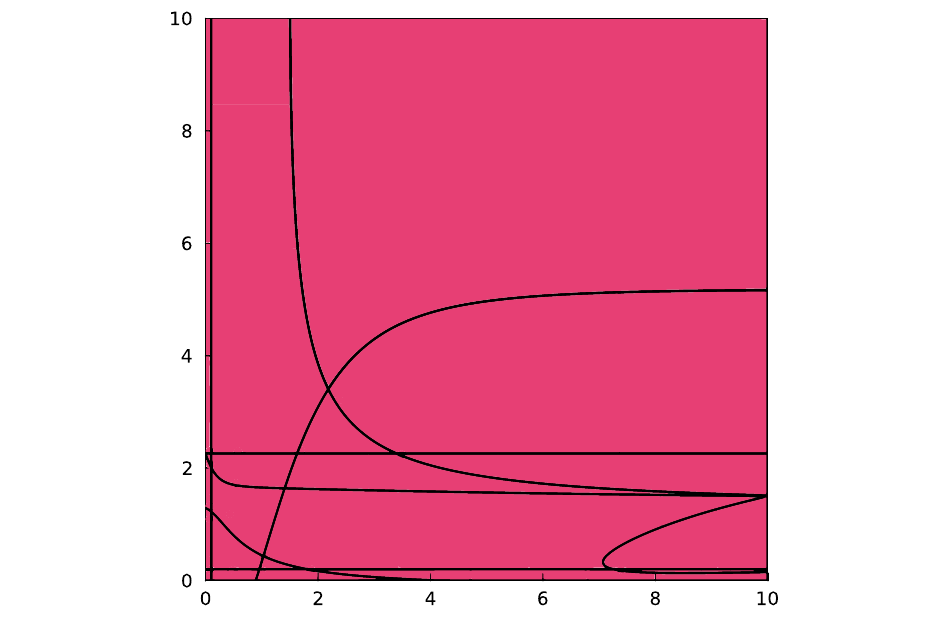}
        \subcaption*{\small (d) $\beta_z = 3.9$}

        \centering
        \includegraphics[width=\linewidth]{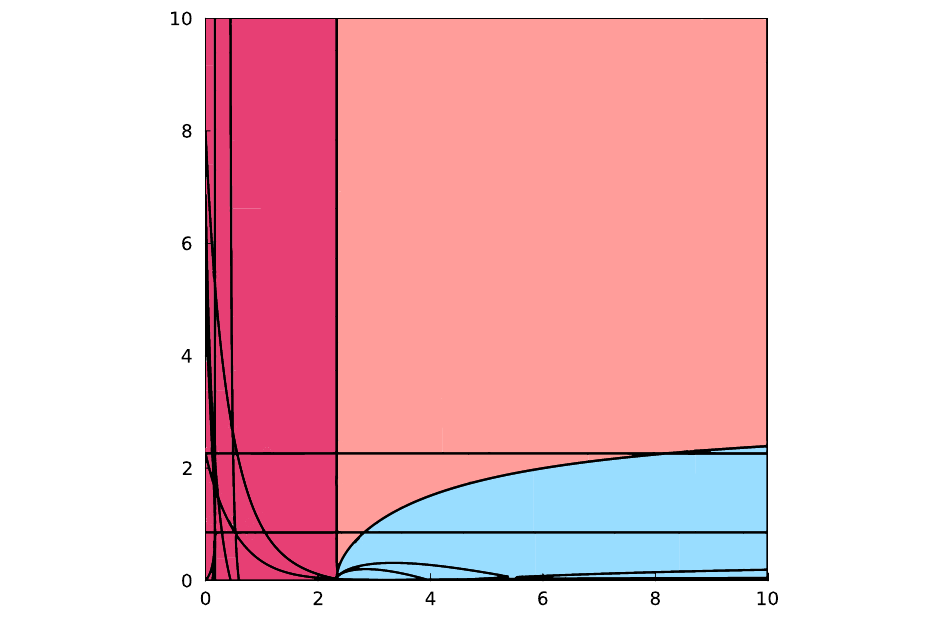}
        \subcaption*{\small (h) $\beta_z = 10$}

    \end{minipage}    
    }
    \caption{Stability landscapes for increasing values of $\beta_z$ with $b = 0.5$. In this case, the intrinsic coral birth rate $b$ is less than the intrinsic coral mortality rate $d=1$. $\tilde{b}$ values are varied along the $x$-axis and $\tilde{d}$ values along the $y$-axis from zero to ten.}
    \label{fig:bx05}
\end{figure}

\subsubsection{\texorpdfstring{Intrinsically non-viable coral population ($b=0.5$)}{Intrinsically non-viable coral population (b=0.5)}}
Figure \ref{fig:bx05} shows the changes in the $b=0.5$ stability landscape as $\beta_z$ is increased from 2.5 to 10. 
In this case, where $b = 0.5 < d = 1$, all stable steady state sets include the total extinction steady state $E_0$. 
Total extinction may be paired with pathogen exclusion, $S_{0,xz} = \left\{E_0,E_{xz}\right\}$, or coexistence, $S_{0,co} = \left\{E_0,E_{co}\right\}$, but never mutualist exclusion. 
In this frame of $\tilde{b}$ and $\tilde{d}$ values, non-extinction steady states were only stable when $\beta_z$ exceeds four (Figure \ref{fig:bx05}, (e)-(h)), that is, when the ability of the mutualistic bacteria to colonize coral was sufficiently high. 
In this case, the non-extinction steady states $E_{xz}$ or $E_{co}$ could only be stable if $\tilde{b}$ was sufficiently large, with this breakpoint decreasing as $\beta_z$ increases. 
If $\tilde{d}$ exceeded approximately 2.5, the coexistence steady state could not be stable (e-h).
Surprisingly, for $\beta_z$ between 4.1 and 6.1, there is also a lower bound for $\tilde{d}$ below which the coexistence steady state $E_{co}$ cannot be stable. 
However, as $\beta_z$ is increased to ten (Figure \ref{fig:bx05} (h)), this lower bound vanishes. 

\subsection{Summary of results}
Increasing the transmissibility of the mutualistic bacteria ($\beta_z$) caused a transformation in the stability landscape, whether the coral host is intrinsically viable ($b>d$) or not ($b<d$). 
When the coral host was intrinsically viable, increasing $\beta_z$ led to a shift in the landscape from one where the coral only steady state $E_x$ was almost always stable (with the possibility of bistability), to one where coral populations always exist and all stable steady states include one of the colonized coral subpopulations. 
In particular, the region of bistability $S_{x,co}$, seen when $\beta_z = 3.9$, abruptly becomes a region of stable coexistence as $\beta_z$ is increased beyond four. 
This suggests that the mutualistic bacteria can facilitate coexistence with the parasitic bacteria through their ability to colonize coral. 
However, stable coexistence always requires that the detrimental effect of the parasitic bacteria ($\tilde{d}$) remains below a certain threshold. 

When the coral population was not intrinsically viable ($b<d$), the possibility of viability emerged when the transmissibility of the mutualistic bacteria ($\beta_z$) exceeded four. 
Still, total extinction was always a stable steady state, no matter the value of $\beta_z$. 
This suggests that if environmental perturbations cause shifts in coral biology leading to their intrinsic inviability, these populations could be preserved if mutualistic bacteria are present. 
Since these regions are bistable, abundances likely must remain high to avoid landing in the stable manifold for total extinction. 
Thus, shocks to populations, such as extreme weather events, could shift stable regimes to ones of total extinction.

\section{Conclusion}\label{sec:Conclusion}

State transitions are ubiquitous in nature, and 
dynamic models are commonly used to understand how perturbations or interventions can induce or prevent such transitions. 
Descriptions of stability landscapes allow for the determination of parameter regimes wherein a desired future state is stable or an unfavored state is unstable. 
Mapping parameter spaces into stability landscapes requires evaluation of a system of polynomial equations and inequalities in several variables, a problem that is often analytically and numerically intractable. 

The routing function method and algorithm presented here provide a straightforward manner of conducting steady-state analysis and numerically determining stability landscapes. 
Routing functions were originally proposed in~\cite{Hong2010} 
for answering connectivity queries and
extended to computing smoothly connected components in~\cite{Cummings2024}.
The proposed extension to analyze stability landscapes 
involves four steps: 
(1) compute the boundary ideal, 
(2)~construct a routing function and compute its critical points, 
(3) track gradient descent/ascent paths from high index critical points to lower index critical points, and 
(4) analyze the possible steady states in each connected component in the complement of the boundary. 
The utility of this method is complemented by the 
availability of computational tools to symbolically compute boundary
ideals and to numerically compute critical points 
and track paths, particularly for the complement of hypersurfaces.

In this work, we illustrated the use of this method through the analysis of the coral-bacteria symbiosis model \citep{Gibbs2024-gx}. 
While this model is relatively simple, describing the whole stability landscape across the five types of equilibria was intractable through standard methods. 
After fixing some parameters, the routing function method allowed us to further elaborate on the conditions in which bacterial colonization supports the persistence of coral populations. 
In particular, we determined that colonization by mutualistic bacteria can facilitate the persistence of coral populations, whether or not the coral population is viable on its own. 
In addition, we found regions of bistability that can be used to determine whether shocks to populations might transition them away from healthy stable states to total extinction.
This method also allowed us to identify a region with no stable steady states for which the system must admit limit cycles or more complex attractors.

For simplicity, we restricted our example to explore the stability landscape induced by certain rate parameters. 
This allowed for a rough qualitative sensitivity analysis of various stable steady sets as these two parameters varied within a given ``frame.''
While this provided a broad theoretical picture of model dynamics, in real applications, the parameters that are treated as variables derive from the available data (or lack thereof) and the particular research question.  Computational efficiency
of the routing function method depends on the feasibility
of symbolically computing boundary ideals
and numerical conditioning of the routing points
and gradient descent/ascent paths.  
Of course, taking two or three parameters as variables is often the best choice when the goal is to produce results that can be interpreted visually. 

While not considered here, the routing function method can be used to address a range of other questions commonly pursued with stability analysis. 
For example, given the existence of bistability or, more generally, $k$-stability, one can treat initial conditions as parameters to determine the stability landscape with respect to the initial conditions. 
Connectivity questions can also be addressed, 
which was the original motivation for developing routing functions.
As demonstrated, this method is particularly well-suited to address common problems in the analysis of ecological models, in particular the description of stability landscapes. 
The adoption of this method by mathematical biologists and ecologists has the potential to expand the 
range of systems that can be analyzed and improve the understanding 
of parameters and their impact on the behavior of the model.

\section*{Acknowledgments}
This material is based upon work supported in part by the National Science Foundation via MPS-Ascend Postdoctoral Research Fellowship Grant No.~MPS-2316455 (KJMD), and and NSF Grant No.~DMS 1945584 (EG),
and NSF Grant No.~CCF 2331400 (JDH),
Simons Foundation 
Grant No.~SFM-00005696 (JDH),
and the Robert and Sara
Lumkins Collegiate Professorship 
(JC and JDH).

\bibliographystyle{abbrv}
\bibliography{references}

\end{document}